\documentclass[twocolumn]{jpsj3}
\begin{document}

\title{\textbf{Fermi Surface Reconstruction inside the Hidden Order Phase of URu$_{2}$Si$_{2}$ Probed by Thermoelectric Measurements} }

\author{
Alexandre~{\sc Pourret}$^1$\thanks{E-mail address: alexandre.pourret@cea.fr}, 
Alexandra~{\sc Palacio-Morales}$^1$,
Steffen~{\sc Kr\"amer}$^2$,
Liam~{\sc Malone}$^3$,
Marc~{\sc Nardone}$^2$,
Dai~{\sc Aoki}$^{1,4}$,
Georg~{\sc Knebel}$^1$\thanks{E-mail address: georg.knebel@cea.fr},
and
Jacques~{\sc Flouquet}$^1$
}

\inst{%
$^1$SPSMS, UMR-E CEA / UJF-Grenoble 1, INAC, Grenoble, F-38054, France\\
$^2$Laboratoire National des Champs Magn\'etiques Intenses, UPR CNRS 3228, CNRS - UJF- UPS - INSA, 25 rue des Martyrs, B.P. 166, 38042 Grenoble cedex 9 France and 143 avenue de Rangueil, 31400 Toulouse, France\\
$^3$H. H. Wills Physics Laboratory, University of Bristol, Tyndall Avenue, BS8 1TL, United Kingdom
$^4$IMR, Tohoku University, Oarai, Ibaraki 311-1313, Japan
}


\abst{
We report thermoelectric measurements of the low carrier heavy fermion compound URu$_{2}$Si$_{2}$ at high fields up to 34T and at low temperatures down to 500mK. The field dependence of the thermoelectric power (TEP) and the Nernst signal shows successive anomalies deep inside the hidden order (HO) phase. The field position of these anomalies correspond to different changes in the Shubnikov-de Haas frequencies and effective masses around 12~T, 17~T, 23~T and 30~T. These results indicate successive reconstructions of the Fermi surface, which imply electronic phase transitions well within the HO phase.
}

\kword{Hidden Order, URu$_2$Si$_2$, Thermoelectric power, Nernst effect}

\maketitle

\section{Introduction}
The nature of the so-called hidden order (HO) state of the low carrier heavy fermion compound URu$_{2}$Si$_{2}$ below the second order phase transition at $T_0 =17.5$~K is still under debate.
The electronic structure changes significantly at the transition from the paramagnetic (PM) to the HO state. Various experimental probes show the gap opening  on the Fermi surface below $T_0$ and most of the charge carriers disappear resulting in a semimetallic state.\cite{Bel2004, Behnia2005} Angular-resolved photoemission spectroscopy observes a narrow dispersive band emerging in the HO state. \cite{Santander-Syro2009, Yoshida2010} The opening of a hybridization gap at $T_0$ has been shown by recent scanning tunneling microscopy experiment \cite{Aynajian2010a, Schmidt2010}. At low temperature, the remaining small numbers of carriers undergo a transition into an unconventional superconducting state at \textit{T$_{sc}$}=1.5 K. \cite{Kasahara2007, Okazaki2008, Yano2008} Recently several exotic order parameters have been proposed for the HO phase including various rank multipole order \cite{Haule2009, Kusunose2012, Ikeda2012, Rau2012}, dynamical symmetry breaking \cite{Elgazzar2009, Oppeneer2010}, spin-nematic states \cite{Fujimoto2011a} or hastatic order \cite{Flint2012}, among others. No higher ordered multipole has been detected by scattering experiments, however dotriacontapole ordering has been proposed from macroscopic experiments, recently.\cite{Okazaki2011,RESSOUCHE2012} 

The HO is affected by external parameters, such as pressure and magnetic field. Above the critical pressure of $P_x \approx 0.5$~GPa, a long range antiferromagnetic ordered state with large moments ($\approx$ 0.4$\mu_{B}$/U) emerges \cite{Amitsuka2007, Hassinger2008}, but no dramatic modification of the Fermi surface is reported between the HO and antiferromagnetic phases \cite{Nakashima2003, Hassinger2010}. In contrast, a strong magnetic field applied along the easy $c$ axis of this tetragonal crystal destroys the HO phase at $H_c \approx 35$~T accompanying a radical reconstruction of the Fermi surface \cite{Jaime2002, Kim2003, Levallois2009, Altarawneh2011}. Above $H_c$ a cascade of several unknown phases are observed just below the metamagnetic field  $H_M \approx 39$~T where a polarized paramagnetic (PPM) metal with large carrier number is recovered. \cite{Kim2003a, Scheerer2012a}


While inside the HO phase ($H < 35$~T for $H\!\!\parallel \!\!c$) no sign of a phase transition can be detected in macroscopic thermodynamic measurements such as the magnetization or the specific heat, topological changes of the Fermi surface (FS) are observed. The first evidence for such FS modifications have been reported by the observation of a new quantum oscillation frequency in the Hall effect slightly below $H^\star = 23$~T \cite{Shishido2009}. Furthermore, a cascade of FS singularities below $H_c$ have been emphasized from the field dependence of the Shubnikov de Haas (SdH) frequencies.\cite{Hassinger2010, Altarawneh2011, Aoki2012} Thermoelectric power (TEP) is known as a very sensitive probe of electronic instabilities and it is a sound approach to study in detail its corresponding response. Recently, TEP has confirmed the singularity at $H^\star = 23$~T and suggested also another anomaly at already $H_m \approx 10$~T.\cite{Malone2011}

In this paper we present a new generation of thermoelectric experiments, extending our previous measurements down to $T \sim 500$~mK and from $H= 28$~T to 34~T, very close to $H_c$. We observe various anomalies in TEP and in the Nernst signal ($N$) inside the HO state. They are marks of a successive Fermi surface evolution deep inside the HO phase. Thus, they establish a clear reference between FS changes and its consequences on transport properties.

\section{Experimental details}

\begin{figure}[ct!]
\includegraphics[width=7.3cm]{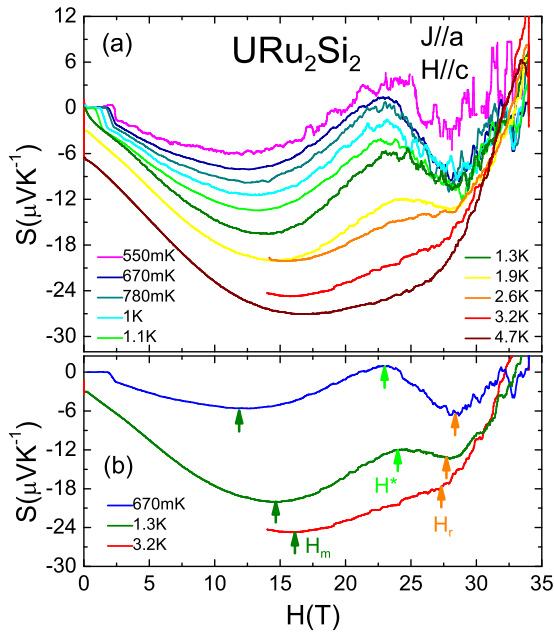}	
\caption{(Color online) (a) Thermoelectric power $S$ of URu$_{2}$Si$_{2}$ for $J\!\!\parallel \!\!a$ and $H \!\!\parallel \!\!c$ as a function of magnetic field for different temperatures. (b) Different anomalies extracted from the TEP and followed as function of temperature: \textit{H$_{m}$} corresponds to the minimum of the thermoelectric power,  \textit{$H^{\ast}$} and  \textit{$H_{r}$} corresponds to the local maximum and minimum of the thermoelectric power at high field.}\label{thermo}
\end{figure}

High-quality single crystals were grown using the Czochralski pulling method in a tetra-arc furnace. Two different samples have been used. Sample 1 $(J\!\!\parallel\!\!a$, $H\!\!\parallel \!\!c$ configuration in the tetragonal crystal structure) had a residual resistance ratio (RRR) of 100, sample 2 ($J\!\! \parallel \!\!H \!\!\parallel \!\!c$ configuration) had RRR of 50. The TEP was measured using the standard ''one heater-two thermometer" setup. Measurements up to 34~T and down to 500~mK were performed in a resistive magnet at the Laboratoire National des Champs Magnetiques Intenses (LNCMI) Grenoble using a recently developed $^{3}$He probe. A significant increase of the noise level had been observed for fields above 20~T which is mainly a consequence of vibrations caused by the water-cooling  of the magnet. Thus, curves are slightly  smoothed for clarity.

\section{Results}

Figure~1(a) shows the TEP as a function of magnetic field up to 34~T for different temperatures. The field is applied along the $c$ axis and the heat current $J$ along the $a$ axis. $S$ is negative revealing an heavy electron band as the dominant heat carrier in the system in the normal state above $H_{c2}$ while the Hall effect is dominated by a light hole Fermi surface.\cite{Kasahara2007} As a function of magnetic field, $S$ shows successive anomalies at low temperature, clearly defined in Fig.~1(b): a rather broad minimum at $H_{m}\sim 11$~T followed by an increase and a local maximum at $H^{\ast} \sim 23$~T and finally another minimum around $H_{r} \sim  30$~T \cite{Aoki2012}.  With increasing temperature both, $H_M$ and $H^\star$ increase in field while $H_r$ decreases and the anomalies at $H^{\ast}$ and $H_{r}$ get apparently smeared out. Above $T\approx 3$~K, well below  the critical temperature corresponding to the critical field $H_{c}$ of the HO phase, the anomalies can no longer be clearly resolved. The field dependence is in good agreement with the previous report.\cite{Malone2011}



\begin{figure}[ct!]
\includegraphics[width=7.3cm]{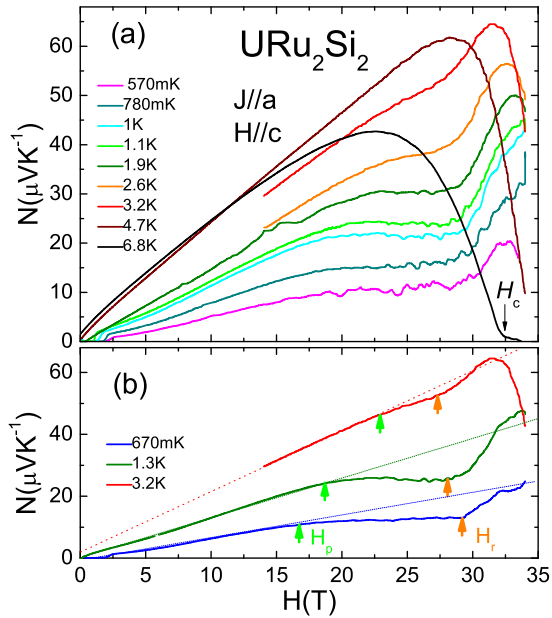}	
\caption{(Color online) (a) Nernst signal of URu$_{2}$Si$_{2}$ for $J\!\!\parallel\!\! a$, $H \!\!\parallel \!\!c$  as a function of magnetic field for different temperatures. $H_c$ indicates the critical field above which the HO is suppressed. (b) The different anomalies extracted from the Nernst signal: $H^{\ast}$ corresponds to the field position where the Nernst signal moves away from a linear field dependence, $H_{r}$ is linked to the abrupt increase of the Nernst signal at high field.}\label{nernst}
\end{figure}

 The Nernst signal ($J\!\! \parallel\!\! a, H\!\! \parallel\!\! c$) measured simultaneously on the same sample as the TEP is shown in Fig.~\ref{nernst}(a). At high temperature, i.e at $T=$ 4.7~K and 6.7~K, we observe the same field dependence as reported in previous measurements performed in pulsed magnetic field \cite{Levallois2009}: the Nernst signal increases almost linearly with the magnetic field, reaches a large maximum and decreases to zero at $H_{c}$ when the HO is suppressed. The low  carrier  number (small Fermi surfaces) in addition to the high mobility of the charge carriers explains the existence of  such a large Nernst signal in the HO phase ($N$ reaches 65 $\mu VK^{-1}$ at 32~T and 3.2~K). In the PM regime, above $H_c$, the Nernst effect is almost zero. This extremely large value of the Nernst signal appears to be an intrinsic property of the HO. However, additional anomalies in the $N (H)$ appear below 3.2~K. $N (H)$ is almost linear up to a field $H_p \sim 17$~T. But for higher fields strong deviations from linearity are observed at low temperature. The field dependence suggests an additive negative contribution to the Nernst signal above $H_p$ getting more and more pronounced on lowering temperature.   For $H > H_{r}$ a sudden increase of the Nernst signal appears and the signal reaches the value from the linear extrapolation from low fields. The positions of the anomalies observed in the field dependence  of the Nernst signal are defined in Fig.~\ref{nernst}(b). Again, these anomalies suggest a change in the relative weight of the dominant heat carriers with increasing  magnetic field up to $H_c$. 

\begin{figure}[t!]	
\includegraphics[width=7.3cm]{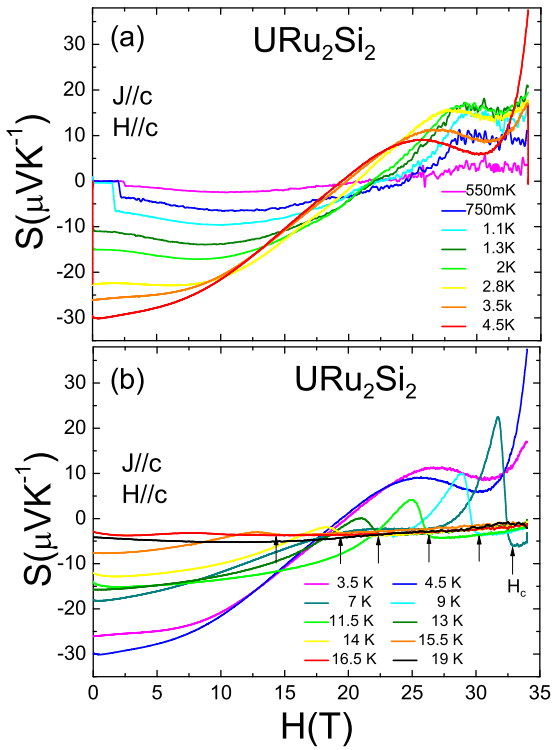}
\caption{\label{thermolong} (Color online)  TEP of URu$_{2}$Si$_{2}$ for $J\!\! \parallel\!\! c$, $H \!\!\parallel\!\!c$ as a function of magnetic field for (a) low temperatures and (b) high temperatures. Above 4.5~K, $S$ presents a peak below the critical field $H_{c}$ (defined by vertical arrows) and in the PM state for $H > H_c$, $S$ is field independent. }
\end{figure}

Previously, strong differences in the field dependence of the transverse and the longitudinal magnetoresistance had been reported\cite{Scheerer2012a}. While the transverse magnetoresistance shows a strong field dependence with a huge maximum at $H_{\rho max}\approx 30$~T \cite{Scheerer2012a}, strongly dependent on the sample quality and dominated by the cyclotron motion of the charge carriers, the longitudinal magnetoresistance does not show such a strong field dependence in the HO state. Thus, we also performed TEP measurements in the longitudinal configuration ($J \parallel \!\!H \parallel\!\! c$), shown in Fig.~\ref{thermolong}. 
In the paramagnetic state for $T > T_0$ the signal is negative and small up to the highest field. Above 4.5~K we clearly see a peak in the TEP just below the critical field $H_{c}$ and in the PM state for $H > H_c$, $S$ is field independent. With lowering the temperature the peak gets more and more pronounced. At low temperature $T<4.5$~K, similar anomalies of the TEP in this longitudinal configuration as in the transverse configuration discussed above can be noticed, but the respective anomalies seem to be shifted to slightly higher fields compared to the transverse configuration. This aspect indicates that the anomalies observed in the different thermoelectric effects are not only governed by scattering but are a direct consequence of successive instabilities  of different Fermi surface pockets at low temperature just below $H_{c}$.

\begin{figure}[ht!]	
\includegraphics[width=9cm]{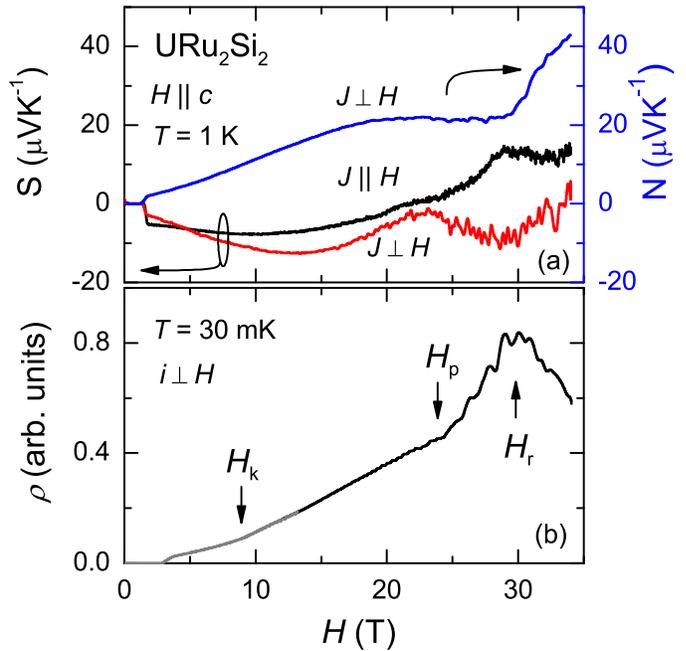}
\caption{\label{magnetoresistivity} (Color online).(a) TEP and Nernst signal obtained at $T = 1$~K. (b) Transverse magnetoresistance for $H\!\! \parallel\!\! c$ at 30~mK: (grey line) from ref.~\citen{Hassinger2010}, (black line) ref.~\citen{Aoki2012}. The arrows indicate the position of anomalies observed in the previous magnetoresistivity experiments.}
\end{figure}

\section{Discussion}

Next, we will compare the thermoelectric response to the field dependent resistivity results.\cite{Hassinger2010, Altarawneh2011, Aoki2012} Fig.~\ref{magnetoresistivity}(a) shows TEP and Nernst signal at 1~K and  (b) the transverse magnetoresistance $\rho (H)$ for $H \!\!\parallel \!\!c$ and $J\!\! \parallel \!\!a$ at $T = 30$~mK as function of field up to 34~T taken from refs.~\citen{Hassinger2010, Aoki2012}. In the magnetoresitance three different anomalies occurs in the raw data, a tiny kink at $H_k \sim 8.5$~T, a second kink with strong enhancement at $H^\star \sim 23$~T and a maximum at $H_r \sim 30$~T. The non-oscillatory part of $\rho (H)$ follows a $H^2$ dependence up to 8~T and in a second regime from 9~T to almost 17~T. A $H^2$ dependence is expected for the transverse magnetoresistance in an compensated metal.\cite{Pippard} Clearly, the field ranges of the thermoelectric anomalies correspond to changes in the magnetoresistance.

\begin{figure}[t!]	
\includegraphics[width=8.3cm]{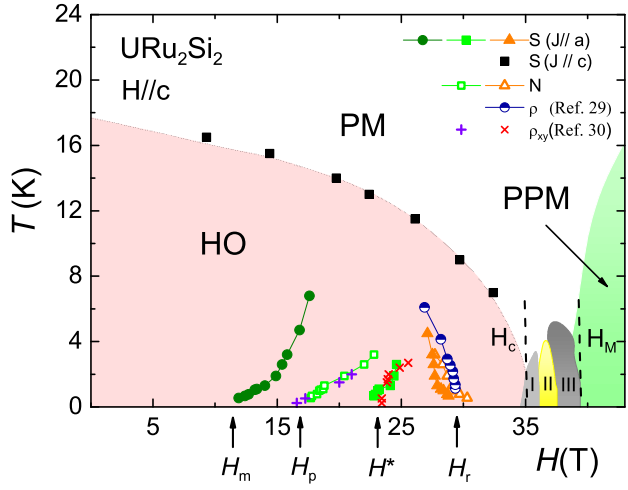}
\caption{\label{diagram} (Color online) $H-T$ Phase diagram of URu$_2$Si$_2$. Several anomalies have been observed inside the HO phase. Full symbols mark anomalies observed in TEP for $H\!\! \parallel\!\! a$ (circles, squares, triangle) and $J \!\!\parallel\!\! c$ (black squares) Anomalies in the Nernst signal are indicated by open symbols. Furthermore, we include anomalies observed in the Hall resistence $\rho_{xy}$ taken from ref.~\citen{Shishido2009} (plus and cross). The indicated anomalies correspond to the maxima in $\rho_{xy}$ $vs H$ of Fig.~1(b) of ref.~\citen{Shishido2009}.  The maximum of resistivity at $H_r$ \cite{Scheerer2012a} (half-filled cicle) are also included. The field position of  the successive changes in the SdH frequencies (\textit{H$_{p}$}, \textit{ H$^{\ast}$}, \textit{H$_{r}$}) are indicated by vertical arrows (see text).\cite{Aoki2012} $H_m$ corresponds to the broad minimum in the TEP. Finally, the different phases which occur between \textit{H$_{c}$} and \textit{H$_{M}$}(I, II, III) and the polarized paramagnetic (PPM) state are indicated. \cite{Scheerer2012a, Kim2003a} This part of the phase diagram was not reached in this report.}
\end{figure}

To summarize the experimental data of successive anomalies inside the hidden order phase, the different anomalies observed in the TEP (full symbol) and in the Nernst signal (open symbol), mentioned above, are reported in the $H-T$ phase diagram  for $H\!\!\parallel \!\!c$, $J\!\!\parallel\!\! a$ in Fig. \ref{diagram}. Using the TEP in the longitudinal configuration, positions of the field independent TEP (black square) delimiting the HO state are reported. Furthermore, we added the anomalies in the Hall resistance shown in Fig.~1 (b) of ref.~\citen{Shishido2009}(half-filled square). As shown in that figure, the Hall effect shows two maxima as function of field, labelled $H_p$ and $H^\star$. (In difference to ref.\citen{Shishido2009} we plot in the phase diagram both, $H_p$ and $H^\star$. We define these characteristic fields as the maxima in the Hall signal $\rho_{xy}$, while in the ref.~\cite{Shishido2009} $H^\star$ is defined as inflection point in $\rho_{xx}$ or in the second derivative $d^2\rho_{xy}/dH^2)$. That's why the curvature of $H^\star (T)$ plotted here is opposite to that in the original paper.) In addition we also plot the maximum of resistivity at $H_r$ taken from in ref.~\citen{Scheerer2012a}) (half-filled circle, in that paper labelled $H_{\rho,max}$. The transport experiments give evidence for several anomalies inside the HO state, while thermodynamic probes do not show any phase line. Furthermore, it should be noticed that all anomalies detected inside the HO state get smeared out for $T>3$~K indicating that they are not conventional phase transitions but related to electronic instabilities of the Fermi surface.

On the basis SdH experiments a successive polarisation of different Fermi surface pockets has been discussed \cite{Hassinger2010, Altarawneh2011, Aoki2012}. Thus, the field position of the successive changes in the SdH frequencies (\textit{H$_{p}$}, \textit{H$^{\ast}$}, $H_{r}$) are indicated by vertical arrows in the phase diagram \cite{Aoki2012}.  The small Fermi surfaces with heavy masses are more easily affected by magnetic field. These changes give strong feedback on the thermoelectric response. $H_m$ corresponds to the broad minimum in the TEP.

\begin{figure}[t!]	
\includegraphics[width=9cm]{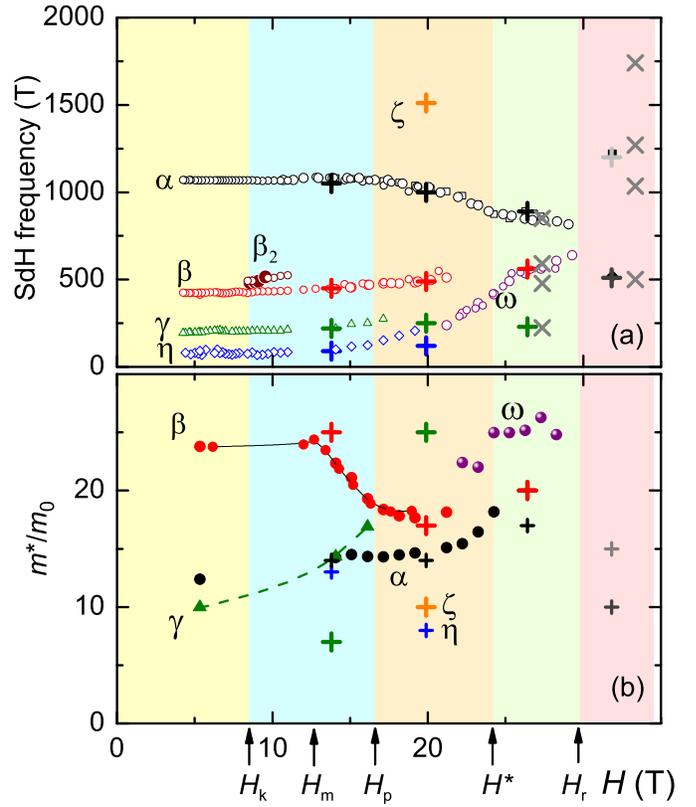}
\caption{\label{SdH} (Color online). (a) Field dependence of the SdH frequencies in the different field ranges. Small open symbols are from ref.~\citen{Hassinger2010} and \citen{Aoki2012}, large pluses are taken from ref.~\citen{Altarawneh2011}, large crosses from ref.~\citen{Jo2007}. The different magnetic field positions \textit{H$_{k}$}, \textit{H$_{m}$}, \textit{H$_{p}$}, \textit{H$^{\ast}$} and \textit{H$_{r}$} are indicated by vertical arrows (see text).}
\end{figure}

Finally we look closer to the previously observed field dependence of the SdH frequencies and the effective cyclotron masses as summarized in Fig.~\ref{SdH}.
In URu$_{2}$Si$_{2}$, several bands have been observed by quantum oscillations \cite{Aoki2012}. Four main quantum oscillation frequencies are observed  at low field $\alpha$, $\beta$, $\gamma$, and $\eta$ with frequencies $F =$ 1.05, 0.42, 0.19, and 0.09 kT, respectively.\cite{Ohkuni1999, Hassinger2010}  The $\alpha$ sheet is most likely attributed to a largest hole Fermi surface centered at the \textit{$\Gamma$} point in the simple tetragonal Brillouin zone, $\beta$ is a four-fold electron Fermi surface located between \textit{$\Gamma$}  and $X$ points, $\gamma$ and $\eta$ are small Fermi surface at the $M$ and $Z$ point, respectively. All quantum oscillation experiments miss an heavy electron Fermi surface which is located at the $M$ point. Possibly, this branch may have  been observed in recent cyclotron resonance experiments.\cite{Tonegawa2012} 

Significant changes of the SdH frequencies and the effective masses $m^\star$ appear under magnetic field as shown in Fig.~\ref{SdH} (a) and (b). It is obvious that in this highly correlated multiband system a modification of one band has strong feedback to all other bands, as at least in the HO state, URu$_{2}$Si$_{2}$ is low carrier compensated metal. As discussed in refs.~\citen{Malone2011} and \citen{Altarawneh2011} the characteristic energies for each band are very low and can be simply estimated by $\Delta_f = \frac{\hbar e F}{m^\star k_B}$. Thus the magnetic fields to fully polarize the bands is rather low and we can estimate that for a field of 20~T all bands except the larger $\alpha$ branch would be polarized.\cite{Altarawneh2011} A first change in the field dependence of the SdH frequencies appears already at relatively small $H_k \approx 8$~T where the magnetoresitance shows a small kink and the spin split of the $\beta$ branch appears \cite{Hassinger2010}. This anomaly in the magnetoresistance is smeared out already at $T\sim 100$~mK. The observed broad minimum in the transverse TEP at $H_m$ appears at $T \sim 500$~mK at slightly higher fields than $H_k$. In the field range of $H_m$ a strong decrease of the effective mass of the $\beta$ branch is observed (see fig.\ref{SdH}(b)). In a multi-band system like URu$_2$Si$_2$, the total Seebeck coefficient $S$ is given by the sum of Seebeck coefficient of each band $S_i$ weighted by its respective conductivity, $S = \sum_i S_i \sigma_i/\sigma$. In difference to conductivity the Seebeck effect can have positive and negative sign, thus the sum of the overall Seebeck effect can be smaller than the contribution of each band. The broad minimum at $H_m$ can be understood as the balance of two different contributions to the TEP. While at low field $H< H_m$ the electron contribution dominates, above $H_m$ a positive hole contribution start to increase as the mass of the electron pocket starts to decrease.    

A smooth decrease of the frequency of the $\alpha$ branch starts above 15~T and reaches nearly $820$~T at $H \sim 29$~T (a reduction of about 20~\%). In this field region the TEP increases up to $H^\star$. This supports the interpretation of the enhancement of the hole contribution in the thermoelectric response, as the mass of observed electron band $\beta$ gets lighter, and vice versa the mass of the hole band $\alpha$ increases slightly. 
Above $H_p\sim 17$~T the appearance of an additional frequency $\epsilon$ has been reported in Hall effect measurements,\cite{Shishido2009} however it has never been reproduced in SdH or de Haas van Alphen experiments.  The effective mass of this $\epsilon$ branch is very low ($m_\epsilon = 2.3$~$m_0$). This field coincides with the anomaly in the Nernst signal, while no signature appears in the Seebeck effect.
Strong field induced modifications of the Fermi surface emerges clearly at $H^\star$ and at $H_{r}$. At $H^\star$ we previously detected the appearance of a new branch named $\omega$ with large cyclotron mass.\cite{Aoki2012} In the field range from $H^\star \approx 23$~T to $H_{r} \approx 30$~T, this new branch coexists at least with the $\alpha$ branch, while an abrupt change of all frequencies appear above $H_{r}$. We observe in the limited field range from $H_{r}$ up to $H_c$ new frequencies in accordance to refs.~\citen{Jo2007, Altarawneh2011}. Remarkably, the sharp anomalies we observed in the thermoelectric response correspond to changes in the Fermi surface topology. In this multiband system magnetic field modifies the balance of the electron and hole contributions to TEP and Nernst effect critically as well as modifications of the Fermi surface topology due to Lifshitz transitions. At $T=0$, these give rise to a diverging TEP,\cite{Malone2011} while the experiments are performed at finite temperature and the divergences are smeared out. 

\section{Summary}

In conclusion, the field dependence of the TEP and the Nernst signal in URu$_{2}$Si$_{2}$ shows successive anomalies deep inside the HO phase. The field position of these anomalies correspond to different changes in the SdH frequencies around 12~T, 23~T and 30~T. These results indicate successive reconstructions of the Fermi surface, which imply electronic phase transitions well within the HO phase. The origin of the phenomena is the Zeeman splitting of the different Fermi sheets. The fact that the TEP anomalies at finite temperature are rather broad appears to be the mark of the strong interplay between the different Fermi sheet. A sharper TEP anomaly has been observed in CeRu$_{2}$Si$_{2}$ at its pseudometamgnetic transition associated with drastic change of the Fermi surface. \cite{Amato89, pfau12} The present work gives a new fact of the link between Fermi surface reconstruction and TEP response in the complex matter of strongly correlated electronic systems.
  
\section*{Acknowledgements}

We thank G.~Scheerer, W.~Knafo and H.~Harima for many useful discussions. This work has been supported by the French ANR (projects PRINCESS, SINUS, CORMAT, DELICE), the ERC (starting grant NewHeavyFermion), the EuromagNET II (EU contract no. 228043) and the Universit\'{e} Grenoble-1 within the Pole SMINGUE.



\end{document}